\documentclass[manuscript]{aastex}
\usepackage{graphicx}

\shorttitle{FRW Universe with interacting sources}
\shortauthors{Dariescu and Dariescu}

\begin{document}

\title{Spatially open Friedmann--Robertson--Walker Universe fueled with interacting sources}

\author{Marina--Aura Dariescu and Ciprian Dariescu }
\affil{Faculty of Physics,
{\it Alexandru Ioan Cuza} University of Ia\c{s}i \\
Bd. Carol I, no. 11, 700506 Ia\c{s}i, Romania}

\email{marina@uaic.ro}

\begin{abstract}
In the present paper, we are analyzing the $k=-1-$FRW Universe with two interacting ideal fluids. The energy densities and the pressures are used to derive the corresponding Equations of State. For a particular form of the interaction term between the two fluids, it turns out that the Universe is filled with a mixture of dust and radiation.
\end{abstract}


\section{Introduction}

In view of Einstein's theory of General Relativity, the accelerating expansion of the Universe was interpreted as
being driven by a  tiny positive cosmological constant.
Even though this assumption is consistent with observational data,
the physical origin of the cosmological constant, its theoretical expectation values which are exceeding the observational
limits and its presence, in all stages of the evolution of the universe, leading to an eternally acceleration, seem
to favor alternative scenarios to nearly flat Universe \citep{cal88,lid99,dva00,sah03}.

Recently, in order to explain  the current CMB anisotropy, supernovae and LSS data, and the age of the Universe, the spatially open Friedmann--Robertson--Walker (FRW) spacetime fueled with different types of matter sources has been seen as a possible candidate
\citep{aur02,gro11}.

Since an open FRW Universe with a sufficiently small curvature can approximate a flat FRW background, in the context of recently proposed nonlinear massive gravity, open FRW Universes driven by arbitrary
matter sources have been thoroughly investigated \citep{gum11}.

Another reason for dealing with negatively curved FRW models is related to the origin and evolution of the
large-scale magnetic field which has been detected not only in galaxies, galaxy clusters and remote protogalactic structures, but also in empty intergalactic space \citep{tav10,and10,ner10,esy11}.

In contrast to the spatially flat FRW cosmological models, where the magnetic field is drastically diluted by the universal expansion, in the case of 
open universes, the cosmological magnetic field can experience a superadiabatic
amplification, irrespective of their matter content and one may assume that this has led to
the strong residual magnetic fields which seeded the galactic dynamo \citep{bar12}.

As for the matter content, besides the ordinary constituents (matter, radiation etc), an exotic fluid called {\it dark energy}, with a large negative pressure, has been added for explaining the expansion history and growth of structure of the Universe \citep{chi05}.
In this respect, the cosmological constant is undoubtedly the simplest and the most appealing candidate and the standard $\Lambda CDM$  (cosmological constant + cold dark matter) model has been remarkably successful in
describing the real Universe, from the matter-dominated phase to the recent accelerated expansion. Nevertheless,
this theory is facing small scale controversies coming from recent observations \citep{veg11, wei13}.
In order to solve this tension, viable alternative models have been formulated, based on various phenomenological non-constant Equations of States (EOS), with
the parameter around the value $w=-1$ \citep{noj15}.

In the present work, we deal with the Friedmann equation with a cosmological constant and two interacting fluids, as sources of a spatially open $(k=-1)-$FRW spacetime. The first fluid is characterized by a dust-type EoS ($w_m =0$) while the second one has a general EoS where the dependence on the Hubble function is manifest, like in the case of viscous fluids. The form of the interaction term is not unique and it is usually postulated from different physical arguments or derived as
a solution of some dynamical equations describing the required properties of
the matter fields \citep{tim09,jam08}.

The case of the dust coupled with dark energy is particularly interesting, leading to an effective EoS and predicting a future evolution of the Universe different from the de Sitter one. The non-constant values of $w$ are supported by measurements of Cosmic Microwave Background Anisotropies, Supernovae luminosity distances and Baryonic Acoustic Oscillations. Recently, by analyzing a large sample of cosmological datasets, a clear indication for $w<-1$, at low redshift, has been found \citep{sai13}.

\section{The $k=-1-$FRW Universe filled with two coupled fluids}

Let us consider the 4-dimensional $k=-1-$FRW metric written as
\begin{equation}
ds^2=a(t)^2 \,\Big[d\chi^2+\sinh^2\chi \,(d\theta^2+\sin^2\theta\,d\varphi^2)\Big]-dt^2,
\end{equation}
where $a(t)=a_0e^{f(t)}$ is the scale factor.

With the Ricci tensor components 
\begin{eqnarray}
& &
R_{\alpha\alpha} \, =  \left[ \frac{\ddot{a}}{a} + 2 \left( \frac{ \dot{a}}{a} \right)^2 - \frac{2}{a^2} \right] \delta_{\alpha \alpha} \; ; \nonumber \\
& &
R_{44} \, =-3 \frac{\ddot{a}}{a}  \, ,
\end{eqnarray}
and the Ricci scalar
\begin{equation}
R=  6 \left[ \frac{\ddot{a}}{a} +  \left( \frac{ \dot{a}}{a} \right)^2 - \frac{1}{a^2}  \right] ,
\end{equation}
the Einstein tensor components, defined by $G_{ab}=R_{ab}-(1/2)R\,g_{ab}$, where $g_{ab}=diag(1,1,1,-1)$, read
\begin{eqnarray}
& &
G_{\alpha\alpha}= \left[ -2  \frac{\ddot{a}}{a} - \left( \frac{ \dot{a}}{a} \right)^2 + \frac{1}{a^2} \right] \delta_{\alpha \alpha} \;
 ; \nonumber \\*
& &
G_{44}= 3 \left[  \left( \frac{ \dot{a}}{a} \right)^2 - \frac{1}{a^2} \right] , 
\end{eqnarray}
where the overdot means the derivative with respect to the cosmic time $t$. 

In view of homogeneity and isotropy of the 
Universe, we assume the energy-momentum tensor of a perfect fluid, characterized by its proper density 
and the pressure $T_{44}=\rho$ and $T_{\alpha\alpha}=P$.
The Einstein's system of equations,
\[
G_{ab} + \eta_{ab} \Lambda =  \kappa T_{ab} \, ,
\]
gets the explicit form
\begin{eqnarray}
& &
 -2 \dot{H} -3 H^2 + \frac{1}{a^2}  +\Lambda= \kappa \,P \; ; \nonumber \\*
& &
3  H^2  - \frac{3}{a^2} -\Lambda= \kappa \, \rho \, ,
\label{eq:expEinstein}
\end{eqnarray}
where $H = \dot{a}/a$ is the Hubble's rate and $\kappa =8\pi\,G/c^2$. 

By combining the two equations in (5), we get the well-known relation
\begin{equation}
\dot{\rho}+3 H \,(\rho+P)=0 \, ,
\end{equation}
where the quantities $\rho$ and $P$ are satisfying the Equation of State $P = w \rho$.

Let us assume the existance of two ideal fluids, with $\rho_m$ and $\rho_d$, with an interaction term between them $Q$, described by the relations
\begin{equation}
\dot{\rho}_m +3 H \, \rho_m=  Q \,  ,
\end{equation}
for the pressureless matter source characterized by $\rho_m$ and $p_m =0$, and
\begin{equation}
\dot{\rho}_d +3 H \,(\rho_d + P_d)= - \; Q \, ,
\end{equation}
for the other source, so that the continuity equation (6) is the summation of (7) and (8).

One of the most common form of the interaction term $Q$ can be found in \citep{bar06},
where the authors are building a model which allows
energy inputs and outflows proportional to the densities of the two fluids, in an expanding FRW universe with zero
spatial curvature.
The total energy being conserved, they have started with the following expression of the
exchanged energy, written in terms of the Hubble parameter and the two energy densities as
\begin{equation}
Q =\left( \gamma \rho_d - \alpha \rho_m \right) H  \,  ,
\end{equation}
where $\alpha$ and $\gamma$ are two positive arbitrary dimensionless constants.

In the followings, we are using the relation (9) and we are postulating an EoS of the form
\begin{equation}
P_d = w \rho_d + \beta H^2 \;  ,
\end{equation}
where, for the moment, $\beta$ (expressed in units of $1/m^2$) is assumed to be positive, $w$ is a constant that will be determined a bit later and $H^2$ is given by the Friedmann's equation
\begin{equation}
H^2 =\frac{\kappa}{3}\left( \rho_m + \rho_d \right) +\frac{\Lambda}{3}+\frac{1}{a^2} \; .
\end{equation} 
Simple corrections to the
net pressure of the form $p = \rho - 3 \xi H$, where $\xi$ is a constant parameter, have been proposed for the FRW Universe filled with {\it stiff fluid} and the FRW Universes have been classified according to the numerical values of the bulk viscous coefficient $\xi$ \citep{mat14}. For a motivation of the necessity of using generalized inhomogeneous EoS fluids, in the context of
singular inflation, we recommend \citep{noj15}.

By switching to the scale function as variable, the equations (7) and (8) become
\begin{equation}
a \rho^{\prime}_m + \left( \alpha + 3 \right) \rho_m=  \gamma \rho_d \,  ,
\end{equation}
\begin{equation}
a \rho^{\prime}_d +3 \left( \rho_d + P_d \right) =  \alpha \rho_m-  \gamma \rho_d  \, ,
\end{equation}
where $\rho^{\prime} = d \rho / da$. These are leading to the following differential equation satisfied by $\rho_m$,
\begin{eqnarray}
& & a^2 \rho^{\prime \prime}_m + a \rho^{\prime}_m  \left[ \alpha + 4 + 3(w+1) + \beta \kappa + \gamma \right]
\nonumber \\*
& & + \left[ 3(w+1)(\alpha  + 3) + 3 \gamma + \beta \kappa \left( \alpha + \gamma + 3 \right) \right] \rho_m 
+  \beta \gamma  \Lambda + \frac{3 \beta \gamma}{a^2} = 0  \nonumber \\*
\end{eqnarray}
while $\rho_d$ is given by (12) as
\begin{equation}
\rho_d = \frac{1}{\gamma} \left[ a \rho^{\prime}_m + \left( \alpha + 3 \right) \rho_m \right]  .
\end{equation}

In order to get a dust matter contribution in the energy density $\rho_m$, one should impose the following relation between the model parameters
\begin{equation}
w = - \, \frac{ (\alpha + \gamma ) \beta \kappa}{3 \alpha} \, ,
\end{equation}
so that the solution of (14) is given by the expression
\begin{equation}
\rho_m \, = \, 
\frac{C}{a^3} - \frac{3 \alpha \beta \gamma}{\alpha ( \alpha +1) + \gamma ( \alpha - \beta \kappa )} \frac{1}{a^2}
- \frac{\alpha \beta \gamma \Lambda}{3 \left[  \alpha ( \alpha +3) + \gamma ( \alpha - \beta \kappa ) \right]}  \, ,
\end{equation}
where the first term corresponds to the dust-type source (with zero pressure), the second term comes as a curvature one, while the last term is proportional to the cosmological constant.
We notice that, for large $a'$s, one is only left with a negative matter energy density, which does actually mean exotic matter, so that, the whole Universe becomes a giant wormhole.

The same type of contributions are in the expression of $\rho_d$, computed with (15), as
\begin{equation}
\rho_d \, = \,
\frac{\alpha C}{\gamma a^3} - \frac{3 \alpha ( \alpha +1) \beta}{\alpha ( \alpha +1) + \gamma ( \alpha - \beta \kappa )} \frac{1}{a^2}
- \frac{\alpha (\alpha +3) \beta \Lambda}{3 \left[  \alpha ( \alpha +3) + \gamma ( \alpha - \beta \kappa ) \right]}  ,
\end{equation}
while the total energy density reads
\begin{equation}
\rho_t \, = \, 
\left( 1 + \frac{\alpha}{\gamma} \right) \frac{C}{a^3} - \frac{3 \alpha \beta ( \alpha + \gamma +1)}{\alpha ( \alpha +1) + \gamma ( \alpha - \beta \kappa )} \frac{1}{a^2}
- \frac{\alpha \beta  (\alpha + \gamma +3) \Lambda}{3 \left[  \alpha ( \alpha +3) + \gamma ( \alpha - \beta \kappa ) \right]}  \; .
\end{equation}

Putting everything together in (10), one is able to derive the pressure
\begin{equation}
P_t = P_d \, = \, 
 \frac{\alpha \beta ( \alpha + \gamma +1)}{\alpha ( \alpha +1) + \gamma ( \alpha - \beta \kappa )} \frac{1}{a^2}
+ \frac{\alpha \beta (\alpha + \gamma +3)\Lambda}{3 \left[  \alpha ( \alpha +3) + \gamma ( \alpha - \beta \kappa ) \right]}  ,
\end{equation}
which is a positive quantity given by the curvature term and the cosmological constant, which keeps open the wormhole throat (pressing against it).

Finally, from the Hubble function written from (10) as
\[
H^2 = \frac{1}{\beta} \left[ P_d - w \rho_d \right] ,
\]
one gets  the relation
\begin{eqnarray}
H^2 & = & \frac{C(\alpha + \gamma) \kappa}{3 \gamma a^3} + \frac{\alpha (\alpha + \gamma +1) - (\alpha +1)(\alpha+\gamma)\beta \kappa}{\alpha ( \alpha + \gamma +1) - \gamma \beta \kappa} \frac{1}{a^2}  \nonumber \\*
& & +
\frac{3 \alpha ( \alpha + \gamma +3) - (\alpha +3)(\alpha + \gamma) \beta \kappa}{9 \left[ \alpha ( \alpha + \gamma +3) - \gamma \beta \kappa \right]}
\Lambda \, ,
\end{eqnarray}
pointing out the special value    
\begin{equation}
\beta \kappa = \frac{\alpha(\alpha + \gamma +1)}{( \alpha +1)(\alpha + \gamma)} \, ,
\end{equation} 
for which $H^2$ gets the simplified expression 
\[
H^2 \, = \, \frac{C(\alpha + \gamma) \kappa}{3 \gamma a^3} 
+
\frac{2 ( \alpha + \gamma ) \left[ 3+\alpha ( \alpha+ \gamma +4) \right]}{9 \left[ 2 \gamma + \alpha  (\alpha + \gamma +1 )(\alpha + \gamma +3) \right]}
\Lambda \, .
\]
By introducing the notations
\begin{equation}
\Lambda_0 =  \frac{2 ( \alpha + \gamma ) \left[ 3+\alpha ( \alpha+ \gamma +4) \right]}{9 \left[ 2 \gamma + \alpha  (\alpha + \gamma +1 )(\alpha + \gamma +3) \right]}
\Lambda \; \; {\rm and} \; \;  D= \frac{C(\alpha + \gamma) \kappa}{3 \gamma}
\end{equation}
one ends up with the equation
\begin{equation}
\left( \frac{\dot{a}}{a} \right)^2 = \Lambda_0  + \frac{D}{a^3} 
\end{equation}
which leads, by integration, to the following scale function 
\begin{equation}
a(t) = \left[ \sqrt{\frac{D}{\Lambda_0}} \sinh \left( \frac{3}{2} \sqrt{\Lambda_0} \, t \right)   
\right]^{2/3} \, .
\end{equation}
The acceleration parameter
\[
q = \frac{\ddot{a}}{a} = \frac{\Lambda_0}{2} \left[ 3 - \coth^2 \left( \frac{3}{2} \sqrt{\Lambda_0} \, t \right)  \right]
\]
turns from negative values, for 
\[
t < \frac{2}{3 \sqrt{\Lambda_0}} {\rm arcsinh} \left( \frac{\sqrt{2}}{2} \right)   \, ,
\]
to positive values afterwards, suggesting an initial decelerating stage followed by an accelerating one.
 
As expected, in view of the previous results, in the very distant future, the expression (25) can be approximated to the de Sitter solution
\[
a \approx \left( \frac{D}{4 \Lambda_0} \right)^{1/3} e^{\sqrt{\Lambda}_0 t}  \; .
\]
For small $t'$s, the scale function is given by
\[
a \sim D^{1/3}  t^{2/3} \, ,
\]
pointing out that this model behaves like a cold-matter-dominated Universe with 
\[
\rho_m \sim \frac{1}{a^3} \, ,
\]
being initiated by a standard {\it Big Bang} process.

The relations (19) and (20), with the condition (22) lead to the following generalized EoS characterizing the Universe with the scale function (25)
\begin{eqnarray}
P_t & = & - \frac{1}{3} \rho_t +   \, \frac{1}{\kappa \Lambda_0 \, a^3} 
+
\frac{2 (\alpha + 1) ( \alpha + \gamma ) ( \alpha+ \gamma +3) \beta }{9 \left[ 2 \gamma + \alpha  (\alpha + \gamma +1 )(\alpha + \gamma +3) \right]} \Lambda \nonumber \\*
& = & - \frac{1}{3} \rho_t + \frac{1}{\kappa} H^2 - \frac{2 \gamma \Lambda}{3 \kappa  \left[ 2 \gamma + \alpha  (\alpha + \gamma +1 )(\alpha + \gamma +3) \right]} \, . 
\end{eqnarray}

\section{The dark energy component}

Let us turn now to the other possible situation, characterized by a negative $\beta$ value. For simplicity, let us just change the $\beta$ sign in (10), i.e.
\begin{equation}
P_d = w \rho_d - \beta H^2 \; , \; \; {\rm with} \; \; \beta >0 \,  .
\end{equation}
Now, $w$ is a positive constant given by
\begin{equation}
w = \frac{ (\alpha + \gamma ) \beta \kappa}{3 \alpha} \, ,
\end{equation}
and the energy densities (17) and (18) are both positive, given by the relations
\begin{equation}
\rho_m \, = \, 
\frac{C}{a^3} + \frac{3 \alpha \beta \gamma}{\alpha ( \alpha +1) + \gamma ( \alpha + \beta \kappa )} \frac{1}{a^2}
+ \frac{\alpha \beta \gamma \Lambda}{3 \left[  \alpha ( \alpha +3) + \gamma ( \alpha + \beta \kappa ) \right]} 
\end{equation}
and
\begin{equation}
\rho_d \, = \,
\frac{\alpha C}{\gamma a^3} + \frac{3 \alpha ( \alpha +1) \beta}{\alpha ( \alpha +1) + \gamma ( \alpha + \beta \kappa )} \frac{1}{a^2}
+ \frac{\alpha (\alpha +3) \beta \Lambda}{3 \left[  \alpha ( \alpha +3) + \gamma ( \alpha + \beta \kappa ) \right]}  .
\end{equation}
These are leading, by summation, to the
(positive) total energy density 
\begin{equation}
\rho_t \, = \, 
\left( 1 + \frac{\alpha}{\gamma} \right) \frac{C}{a^3} + \frac{3 \alpha \beta ( \alpha + \gamma +1)}{\alpha ( \alpha +1) + \gamma ( \alpha + \beta \kappa )} \frac{1}{a^2}
+ \frac{\alpha \beta  (\alpha + \gamma +3) \Lambda}{3 \left[  \alpha ( \alpha +3) + \gamma ( \alpha + \beta \kappa ) \right]}  ,
\end{equation}
where the constant term can be identified with  the
cosmological density $\rho_{\Lambda}$.
The pressure, computed with (27), corresponds to a dark energy component,
\begin{equation}
P_t =  P_d \, = \, 
- \frac{\alpha \beta ( \alpha + \gamma +1)}{\alpha ( \alpha +1) + \gamma ( \alpha + \beta \kappa )} \frac{1}{a^2}
- \frac{\alpha \beta (\alpha + \gamma +3)\Lambda}{3 \left[  \alpha ( \alpha +3) + \gamma ( \alpha + \beta \kappa ) \right]}  \, ,
\end{equation}
coming from the curvature and the cosmological constant contributions.

The Hubble function coming from (27), i.e.
\begin{eqnarray}
H^2 & = & \left( \frac{\dot{a}}{a} \right)^2 = \frac{1}{\beta} \left[ w \rho_d - P_d \right] \nonumber \\*
& = & \frac{C(\alpha + \gamma) \kappa}{3 \gamma a^3} + \frac{\alpha (\alpha + \gamma +1) + (\alpha +1)(\alpha+\gamma)\beta \kappa}{\alpha ( \alpha +1) + \gamma ( \alpha + \beta \kappa )} \frac{1}{a^2}  \nonumber \\*
& & +
\frac{3 \alpha ( \alpha + \gamma +3) + (\alpha +3)(\alpha + \gamma) \beta \kappa}{9 \left[\alpha ( \alpha +3) + \gamma ( \alpha + \beta \kappa ) \right]}
\Lambda \equiv \frac{D}{a^3} + \frac{F}{a^2}  +  \Lambda_0 \,  ,
\end{eqnarray}
leads, by integration, to an elliptic function from which the dependence $a(t)$ is impossible to derive analytically.

However, in the particular case of zero cosmological constant, the relation (33) gets the simple form
\[
\frac{da}{\sqrt{F+ \frac{D}{a}}} = dt \, ,
\]
leading to the transcendental equation
\begin{equation}
t-t_* = \frac{1}{F} \sqrt{a(aF+D)}- \frac{D}{F \sqrt{F}}  {\rm arcsinh} \sqrt{\frac{aF}{D}} \, ,
\end{equation}
with $t_*$ an integration constant corresponding to a special universal (cosmic) moment. 
Based on the substitution 
\[
{\rm arcsinh}\sqrt{\frac{aF}{D}} = \chi \, ,
\]
we can derive from (34) the following representation
\begin{eqnarray}
& & a(\chi)  = \frac{D}{F} \, \sinh^2(\chi) \nonumber \\*
& &
t-t_*  =\frac{D}{2F \sqrt{F}}\left[ \sinh(2\chi) -2\chi \right] .
\end{eqnarray}

From the system above, we notice that for $\chi_*=0$ and $t=t_*$ we get $a(\chi_*)=0$, pointing out the singular cosmic event, at $t_*$. Also, for small $\chi$, when the density of matter term dominates the curvature term, the behavior is $a\sim t^{2/3}$, as in the previous case. On the other hand, as the time evolves, the curvature of space starts to dominate the matter density, which is more and more diluted and we are recovering the \emph{Milne model} of the Universe, with $a\sim t$.

The equation of state parameter $w_d = P_d / \rho_d$, computed with (32) and (30) for $\Lambda =0$, is a function of $\chi$, decreasing from zero, at $\chi \to 0$, and tending to the negative constant value
\begin{equation}
w_* = - \frac{(\alpha + \gamma +3)(3 + 2 \beta \kappa)}{3(\alpha +3)}= - \left[ 1 +  \frac{\gamma}{\alpha + 3}+ \frac{2 \beta \kappa (\alpha+\gamma+3) }{3(\alpha+3)} \right] ,
 \end{equation}
in the far future.

\section{A particular form of the interaction term}

As in \citep{bah16}, let us consider the particular case $\gamma =0$ so that the interaction term is $Q = - \alpha H \rho_m$.
Such a choice for an open Universe has been proved to be consistent with the observed Universe expansion epochs, i.e. first deceleration and then
acceleration.

The same expression of $Q$ has been used in the analysis of generalized ghost
pilgrim dark energy in non-flat
FRW Universe \citep{jaw14}, the negative sign meaning that the CDM decays into DE.

The particular form of the system (12, 13), i.e.
\begin{equation}
a \rho^{\prime}_m + ( \alpha + 3 ) \rho_m=  0 \,  ,
\end{equation}
\begin{equation}
a \rho^{\prime}_d +3 \left( \rho_d + P_d \right) =  \alpha  \rho_m  \, ,
\end{equation}
with $P_d$ given in (10), is satisfied by the energy densities
\begin{equation}
\rho_m = \frac{C_1}{a^{3+\alpha}} \, ,
\end{equation}
and
\begin{equation}
\rho_d = \frac{C_2}{a^{3w+\beta \kappa+3}} + \frac{( \alpha - \beta \kappa)C_1}{(3w+\beta \kappa - \alpha)a^{3+ \alpha}} - \frac{3 \beta}{(3w+ \beta \kappa+1)a^2} -\frac{\beta \Lambda}{3w+\beta \kappa +3} \, ,
\end{equation}
so that the total energy density reads
\begin{equation}
\rho_t = \frac{C_2}{a^{3w+\beta \kappa+3}} + \frac{3wC_1}{(3w+\beta \kappa -\alpha)a^{3+\alpha}} - \frac{3 \beta}{(3w+ \beta \kappa+1)a^2} -\frac{\beta \Lambda}{3w+\beta \kappa +3} 
\end{equation}
and the total pressure is given by
\begin{equation}
P_t = \frac{C_2 (3w + \beta \kappa)}{3 a^{3w+\beta \kappa+3}} + \frac{\alpha wC_1}{(3w+\beta \kappa -\alpha)a^{3+\alpha}} + \frac{\beta}{(3w+ \beta \kappa+1)a^2} + \frac{\beta \Lambda}{3w+\beta \kappa +3}  \; .
\end{equation}

One may notice that, for the particular value $\alpha =1$, the energy density $\rho_m$ is corresponding to radiation, while the relations (40) and (41) become
\begin{equation}
\rho_d = \frac{C_2}{a^{3w+\beta \kappa+3}} + \frac{(1- \beta \kappa)C_1}{(3w+\beta \kappa -1)a^4} - \frac{3 \beta}{(3w+ \beta \kappa+1)a^2} -\frac{\beta \Lambda}{3w+\beta \kappa +3} \, ,
\end{equation}
and 
\begin{equation}
\rho_t = \frac{C_2}{a^{3w+\beta \kappa+3}} + \frac{3wC_1}{(3w+\beta \kappa -1)a^4} - \frac{3 \beta}{(3w+ \beta \kappa+1)a^2} -\frac{\beta \Lambda}{3w+\beta \kappa +3} \, .
\end{equation}

For the relation (16) between the model parameters, which in this particular case is
$3w + \beta \kappa =0$, one deals with an Universe filled with a mixed source made of dust and radiation, with the total energy
\begin{equation}
\rho_t = \frac{C_2}{a^3} + \frac{\beta \kappa C_1}{a^4} - \frac{3 \beta}{a^2} -\frac{\beta \Lambda}{3}  \; .
\end{equation}
The corresponding pressure,
\begin{equation}
P_t = P_d = \beta \left[ \frac{\kappa C_1}{3 a^4} + \frac{1}{a^2} + \frac{\Lambda}{3} \right]    ,
\end{equation}
is negative for negative values of $\beta$, pointing out a dark energy component, as in the general analysis developed in section 3.

The other particular case with $3w + \beta \kappa = 3$ is
characterizing an Universe with stiff matter and radiation with the total energy density
\begin{equation}
\rho_t = \frac{C_2}{a^6} + \frac{(3 - \beta \kappa)C_1}{2 a^4} - \frac{3 \beta}{4 a^2} -\frac{\beta \Lambda}{6}
\end{equation}
and pressure
\begin{equation}
P_t = P_d = \frac{C_2}{a^6} + \frac{(3 - \beta \kappa)C_1}{6 a^4} + \frac{\beta}{4 a^2} + \frac{\beta \Lambda}{6} \, .
\end{equation}
For negative $\beta$, the energy density is always positive and the EoS parameter $w_t = P_t / \rho_t$ is decreasing from the maximum value $w_t =1$, for $a \to 0$ to $w_t =-1$, in the far future, gaining all the values inbetween.

\section{Conclusions}

In the present paper, we have considered a $k=-1-$FRW Universe filled with two interacting ideal fluids, characterized by $P_m=0$ and a general EoS of the form (10) respectively, and we have derived the total energy density, the pressure and the corresponding scale function. 

The main constraint on the model parameters, (16), is generated by imposing the existence of a pressureless matter term in the energy density (17), besides the curvature and the cosmological constant contributions.

One may notice that, with the special value of the coupling parameter (22), the simplified relation (24) can be written, in terms of the redshift, as
\[
H^2 = \frac{D}{a_0^3} ( 1+z)^3 + \Lambda_0 \, .
\]
This is of the form 
\[
H^2 = H_0^2 \left[  \Omega_m^0  (1+z)^3 +  \Omega_{\Lambda} \right] ,
\]
corresponding to the flat $\Lambda CDM$ model, with 
\[
D = a_0^3 H_0^2 \Omega_m^0 \; , \; \;
\Lambda_0 = H_0^2 \Omega_{\Lambda} \; , \; \; \frac{D}{a_0^3}  + \Lambda_0 \, = H_0^2 \,  ,
\]
where the index ^^ ^^ 0'' is for the present time values. In order to determine the values of the model's parameters $D$ and $\Lambda_0$, one may use the current fractional density of matter estimated from Nine-Year WMAP Observations,
$\Omega_m^0 = 0.1369 \cdot h^{-2} = 0.276$, for the present Hubble parameter $h = H_0 / 100 / [km \cdot sec^{-1} \cdot Mpc^{-1} ] = 0.705$ \citep{wma13}. 

Even though, in the last two decades, the $\Lambda CDM$ model has been remarkably successful in describing the large-scale structure of the Universe,  the predictions at small scales are controversial and disagree with recent observations \citep{bos15}.
Among the attempts to relieve the tension, we mention the ones which are changing the nature of spacetime to a non-flat background \citep{kum15} or the nature of the dark matter, by adding a hot component \citep{jeo14} or
a viscous one \citep{vel14}. 

Nevertheless, for values of $\beta$ which are not fulfilling the condition (22), the form of the Hubble function is
\begin{eqnarray}
H^2 & = & \frac{D}{a_0^3} ( 1+z)^3 +  \frac{[\alpha (\alpha + \gamma +1) - (\alpha +1)(\alpha+\gamma)\beta \kappa]}{[ \alpha ( \alpha + \gamma +1) - \gamma \beta \kappa] a_0^2} (1+z)^2 + \Lambda_0 \nonumber \\* 
& = &
\frac{D}{a_0^3} ( 1+z)^3 + f(z) + \Lambda_0 \, , \nonumber
\end{eqnarray}
and the lower values of $H$ at higher redshifts can be achieved when the dynamically evolving term $f(z)$ is negative.
Among the numerous forms of this screening term,
the curvature one has been considered as a natural and plausible choice  \citep{kum15}, to satisfy the results from BOSS experiment \citep{bos15}.

On the other hand, the EoS (26) is similar to the one of a inhomogeneous viscous dark fluid coupled with dark matter in the FRW Universe, $p  = w \rho -3 H \xi ( H , t)$,
where $\xi (H,t)$ is the bulk viscosity. In the simplest case, $\xi$ is taken as a positive constant, $\xi = \xi_0$, while in other theories, it is proportional to $H$ as
$\xi = 3 \tau H$, with $\tau$ a positive constant \citep{eli14}.

For negative values of the model parameter $\beta$, it turns out that the dark energy pressure (32) is made of two negative contributions coming from the curvature and the cosmological constant. 

In the particular case of zero cosmological constant,
the Universe evolves from the singular cosmic event, at $t_*$, to the Milne stage, in the far future.

The form of the EoS parameter (36) is sustained by a number of scalar
field models which have proposed a dynamical (time-dependent) $w$ parameter, such as quintessence (with $w \geq -1$), phantom ($w$ lies below -1) \citep{cal02,cal03} and quintom matter which crosses over
the cosmological constant boundary $w=-1$ \citep{fen05,cai10}.

In the final part of the paper, we have concentrated on the special case of one-direction outflow, from the postulated dust-type component characterized by $\rho_m$ and $p_m =0$ to the second component with the general EoS (10). Quite intriguing, if one assumes  $\alpha =1$, the first component is behaving radiation-like, until it starts to earn energy, according to (9).

\end{document}